\documentclass[a4paper]{jpconf}

\usepackage{graphicx}
\usepackage{etex, xy,color}
\usepackage{mathrsfs}
\usepackage{amsmath,amsbsy,amssymb}
\usepackage{citesort}

\xyoption{all}

\newcommand{\ep}{\varepsilon}
\let\set\mathbb
\newcommand{\vect}[1]{\vec{#1}}

\newcommand{\Mfrac}[2]{\frac{\text{\tiny$\displaystyle#1$}}{\text{\tiny$#2$}}}
\newcommand{\pisiSE}{$\Pi\Sigma^*$}
\newcommand{\KK}{\mathbb{K}}
\newcommand{\NN}{\mathbb{N}}
\newcommand{\FF}{\mathbb{F}}
\newcommand{\QQ}{\mathbb{Q}}
\newcommand{\dfield}[2]{({#1},{#2})}
\newcommand{\SigmaP}{\texttt{Sigma}}

\newcommand{\fct}[3]{{#1:#2 \to #3}}
\newcommand{\myShift}{{\mathscr{S}}}

\newcommand{\MyFrame}[1]{
\hspace*{-0.08cm}\fcolorbox[rgb]{0,0,0}{1,1,0.65}{
\begin{minipage}{15.5cm}#1\end{minipage}}
}

\newcounter{linectr}
\newenvironment{myEnumerate}{\begin{list}{(\arabic{linectr})}{\usecounter{linectr}
\labelwidth1ex\itemsep0ex\labelsep1ex\leftmargin2ex\parskip0.0cm\topskip0cm\partopsep0cm
\listparindent0ex}}{\end{list}}

\newcounter{Cmmacnt}
\def\restartmma{\setcounter{Cmmacnt}{0}}
\restartmma \catcode`|=\active
\def|#1|{\mathrm{#1}}
\catcode`|=12
\newenvironment{Cmma}{
 \par\smallskip
 \catcode`|=\active
 \parskip=0pt\parindent=0pt 
 \footnotesize
 \def\CIn##1\\{%
   \def\linebreak{\hfill\break\null\qquad}%
   \refstepcounter{Cmmacnt}
   \hangindent=2.5em\hangafter=0
   \leavevmode
   \llap{\tiny\sffamily In[\arabic{Cmmacnt}]:=\kern.5em}%
   \mathversion{bold}\scriptsize$\tt\bf\displaystyle##1$\normalsize
   \mathversion{normal}
 }%
 \def\CPrint##1\\{%
   \def\linebreak{$\hfill\break\null\hfill$}%
   \kern\abovedisplayskip\par
   \hangindent=2.5em\hangafter=0
   \leavevmode
   \llap{}
   \scriptsize$\displaystyle\tt##1$\normalsize\hfill\null\par
   \kern\belowdisplayskip
 }%
 \def\COut##1\\{%
   \def\linebreak{$\hfill\break\null\hfill$}%
   \kern\abovedisplayskip\par
   \hangindent=2.5em\hangafter=0
   \leavevmode
   \llap{\tiny\sffamily Out[\arabic{Cmmacnt}]=\kern.5em}
   \scriptsize$\displaystyle\tt##1$\normalsize\hfill\null\par
   \kern\belowdisplayskip\vspace*{-0.3cm}
 }%
 \def\CWarning##1##2\\{%
   \def\linebreak{\hfill\break}%
   \hangindent=2.5em\hangafter=0
   \leavevmode
   {\scriptsize##1 : ##2}\par}%
}{%
 \par\smallskip
}

\newcommand{\CmyIn}[1]{{\small\sffamily In[#1]}}
\newcommand{\CmyOut}[1]{{\small\sffamily Out[#1]}}
\def\CMLabel#1{{\refstepcounter{Cmmacnt}\label{#1}}\addtocounter{Cmmacnt}{-1}}

\begin{document}
\title{Modern Summation Methods for Loop Integrals in Quantum Field Theory: The Packages Sigma, EvaluateMultiSums and SumProduction}

\author{C. Schneider}

\address{Research Institute for Symbolic Computation (RISC)\\
Johannes Kepler University Linz\\ 
Altenbergerstr. 69,  4040 Linz, Austria}

\ead{Carsten.Schneider@risc.jku.at}

\begin{abstract}
A large class of Feynman integrals, like e.g., two-point parameter integrals with at most one mass and containing local operator
insertions, can be transformed to multi-sums
over hypergeometric expressions.
In this survey article we present a difference field approach for symbolic summation
that enables one to simplify such definite nested sums to indefinite nested sums. In particular, the simplification is given --if possible-- in terms of harmonic sums, generalized harmonic sums, cyclotomic harmonic sums or binomial sums.
Special emphasis is put on the developed packages \texttt{Sigma}, \texttt{EvaluateMultiSums} and \texttt{SumProduction} that assist in the task to perform these simplifications completely automatically for huge input expressions. 
\end{abstract}

\section{Introduction}\label{Sec:Intro}

This survey article aims at giving an overview of the available summation tools and packages in the setting of difference fields and aims at providing the basic insight how they work and how they can be applied to Feynman integrals.\\
The symbolic summation approach under consideration started with Karr's telescoping algo\-rithm in \pisiSE-fields~\cite{Karr:81}. There one can treat indefinite nested product-sum expressions. More precisely, one can search for optimal sum representations for these expressions such that the arising sums have minimal nesting depth~\cite{Schneider:05f,Schneider:08c,Schneider:10b}, the polynomial expressions in the sums have minimal degrees~\cite{Schneider:04a,Schneider:07d} and such that the used sums are algebraically independent~\cite{Schneider:10c,Schneider:13a}. In particular, one can discover and prove definite sums over such expressions by exploiting the summation paradigms of parameterized telescoping and recurrence finding~\cite{Schneider:01,Schneider:04c}. In a nutshell, this toolkit implemented in the summation package \SigmaP~\cite{Schneider:07a} can be considered as a generalization of the well known hypergeometric summation machinery~\cite{Zeilberger:91,Petkov:92,AequalB}.
Meanwhile these difference field tools have been applied successfully, e.g., in combinatorics, number theory or numerics; for instance, we refer to~\cite{Schneider:03,APS:05,Schneider:06c,Schneider:09a}.

In the last 7 years I have pushed forward these summation technologies and have developed new packages~\cite{Schneider:13a,ABRS:12,Schneider:12a,Schneider:10d} to carry out challenging calculations in particle physics. 
In cooperation with J. Bl\"umlein (DESY, Zeuthen), e.g.,
Feynman integrals with at most one mass have been considered. They are defined in $D$-dimensional Minkowski space with one time- and $(D-1)$
Euclidean space dimensions with $\varepsilon = D - 4\in {\mathbb R}$ where
$|\varepsilon| \ll 1$.
The integrals depend on a discrete Mellin parameter $n$ which comes from local operator insertions. 
As worked out in~\cite{Bluemlein:09,BKSF:12} these integrals can be transformed
to integrals of the form
\begin{eqnarray}
\label{Equ:HypInt}
{\cal I}(\ep,n) = C(\ep, n, M) \int_0^1 dx_1 \ldots \int_0^1 dx_m
\frac{\sum_{i=1}^k \prod_{{l}=1}^{r_i}
[P_{i,l}(x_1,\dots,x_m)]^{\alpha_{i,l}(\varepsilon,n)}}{[Q(x_1,\dots,x_m)]^{\beta(\varepsilon)}}~,
\end{eqnarray}
with $k\in\set N=\{0,1,2,\dots\}$, $r_1,\dots,r_k\in\set N$ and the following ingredients. 
$C(\ep, n, M)$ is a factor, which depends on the dimensional parameter $\ep$,
the integer parameter $n$ and the mass $M$ (being possible $0$).
The $P_i(x_1,\dots,x_m)$ and  $Q(x_1,\dots,x_m)$ are polynomials in the integration variables $x_i$. Further, the exponent $\beta(\ep)$ is a rational function in $\ep$, i.e., $\beta(\ep)\in\set Q(\ep)$, and similarly 
$\alpha_{i,l}(\ep,n) = n_{i,l} n + \overline{\alpha}_{i,l}$ with $n_{i,l} \in \{0,1\}$ and $\overline{\alpha}_{i,l}\in\set Q(\ep)$. For integrals without local operator insertions  see also \cite{BW:10,Weinzierl:13}.\\
Then given such an integral we seek for a Laurent series expansion for some $t\in\set Z$:
\begin{equation}\label{Equ:LaurentExp}
{\cal I}(\ep,n)=I_{t}(n)\ep^t+I_{t+1}(n)\ep^{t+1}+I_{t+2}(n)\ep^{t+2}+\dots
\end{equation}
More precisely, the crucial task is to determine the first coefficients $I_i(n)$ in closed form.

\medskip

\noindent\textit{Remark.} There are computer algebra tools~\cite{ABRS:12} available to solve this problem for integrals of the form~\eqref{Equ:HypInt}. Namely, we can calculate a recurrence in $n$ for~\eqref{Equ:HypInt} using Ablinger's \texttt{MultiIntegrate} package~\cite{Ablinger:12}, an optimized and refined
version of the Almkvist--Zeilberger algorithm~\cite{AZ:06}. Given such a recurrence, we can calculate the desired $\ep$-expansion by algorithms~\cite{BKSF:12} given in \SigmaP. We remark that the method of hyperlogarithms~\cite{Brown:09} can be also used to evaluate integrals of the form~\eqref{Equ:HypInt} for specific values $n\in\set N$ if one can set $\ep=0$. An adaption of this method for symbolic $n$ has been described and applied to massive 3-loop ladder graphs in~\cite{ABHKSW:12}.

\medskip

For some types of integrals these approaches work nicely. However, for many cases it was more suitable to proceed as follows. Using the method in~\cite{BKSF:12} the given integral~\eqref{Equ:HypInt} can be transformed to proper hypergeometric multi-sums of the format\footnote{For convenience, we assume that the summand is written terms of the Gamma function $\Gamma(x)$. Later, also Pochhammer symbols or binomial coefficients are used which can be rewritten in terms of Gamma-functions.}
\begin{equation}\label{Eq:GenericMultiSum}
{\cal S}(\ep,n) = \sum_{n_1=1}^\infty ... \sum_{n_r=1}^\infty
\sum_{k_1=1}^{L_1(n)} ... \sum_{k_v=1}^{L_v(n,k_1, ..., k_{v-1})}
\sum_{i=1}^l C_i(\ep, n, M)
\frac{\Gamma(t_{1,i}) \ldots \Gamma(t_{v',i})}	
     {\Gamma(t_{v'+1,i}) \ldots \Gamma(t_{w',i})}.
\end{equation}
Here the upper bounds $L_1(n),\dots,L_{v}(n,k_1,\dots,k_{v-1})$ are $\infty$ or integer linear expressions in
the dependent parameters (i.e., linear combinations of the variables over the integers), and $t_{l,i}$ are linear combinations of the $n_1,\dots,n_r,k_1,\dots,k_v,\ep$ over $\set Q$.

\medskip

\noindent\textit{Remark.} At this level, one can apply similar techniques as for the integral representation~\eqref{Equ:HypInt}. Namely, as carried out in~\cite{BKSF:12} we can calculate recurrence relations for such multi-sums~\eqref{Eq:GenericMultiSum} by using either the WZ-approach~\cite{Wilf:92} and efficient and refined algorithms developed in~\cite{Wegschaider}. Similarly, we can use a common framework~\cite{Schneider:05d} within the summation package \SigmaP\ that combines difference field~\cite{Schneider:05a} and holonomic summation techniques~\cite{Chyzak:00} to compute recurrences for the given sum~\eqref{Eq:GenericMultiSum}. Then given such a recurrence, one can use again \SigmaP's recurrence solver to calculate the coefficients of the Laurent-series expansion~\eqref{Equ:LaurentExp}.

\medskip

Finding recurrence relations for the integrals of the form~\eqref{Equ:HypInt} and sums of the form~\eqref{Eq:GenericMultiSum} involving the $\ep$-parameter is a rather tough problem. However, we can get rid of the parameter $\ep$, if the sums~\eqref{Eq:GenericMultiSum} are uniformal convergent: First one expands the summand of~\eqref{Eq:GenericMultiSum}, say
\begin{equation}\label{Equ:SummandExpand}
F(n,n_1,\dots,n_r,k_1,\dots,k_v)=F_t(n,n_1,\dots,k_v)\ep^t+F_{t+1}(n,n_1,\dots,k_v)\ep^{t+1}+\dots
\end{equation}
with $t\in\set Z$
by using formulas such as~\cite{BKSF:12}
\begin{equation*}
\Gamma(k+1+\bar{\ep}) = \frac{\Gamma(k) \Gamma(1+\bar{\ep})}{B(k,1+\bar{\ep})}\text{ and }B(k, 1 + \bar{\ep}) = \frac{1}{k}\exp\left(\sum_{i=1}^\infty \frac{(-\bar{\ep})^i}{i} S_i(k)\right)
= \frac{1}{k}\sum_{i=0}^\infty (-\bar{\ep})^i S_{\underbrace{\mbox{\scriptsize 1, \ldots
,1}}_{\mbox{\scriptsize
$i$}}}(k)
\end{equation*}
with $\bar{\ep} = r \ep$ for some $r\in\set Q$.
Here 
$B(x,y)=\Gamma(x)\Gamma(y)/\Gamma(x+y)$ denotes the Beta-function and $S_{\mbox{\scriptsize 1, \ldots
,1}}(n)$ is a special instance of the harmonic sums~\cite{Vermaseren:99,Bluemlein:99} defined by
\begin{equation}\label{Equ:HarmonicSums}
S_{c_1,\dots,c_r}(k)=
\sum_{i_1=1}^k\frac{\text{\small$\text{sign}(c_1)^{i_1}$}}{i_1^{|c_1|}}\sum_{i_2=1}^{i_1}\frac{\text{\small$\text{sign}(c_2)^{i_2}$}}{i_2^{|c_2|}}\dots
\sum_{i_r=1}^{i_{r-1}}\frac{\text{\small$\text{sign}(c_r)^{i_r}$}}{i_r^{|c_r|}}
\end{equation}
with $c_1,\dots,c_r$ being nonzero integers. If there are no poles\footnote{If there are poles, these extra evaluations are treated separately by first plugging the values into $f$ and by expanding this expression afterwards.} withing the summation range, one can apply the summation signs to each of the coefficients in~\eqref{Equ:SummandExpand}. I.e., the $i$th coefficient of the $\ep$-expansion~\eqref{Equ:LaurentExp} of~\eqref{Eq:GenericMultiSum} can be written in the form
\begin{equation}\label{Equ:MultiSumsOverIndefNested}
I_i(n)=\sum_{n_1=1}^\infty ... \sum_{n_r=1}^\infty
\sum_{k_1=1}^{L_1(n)} ... \sum_{k_v=1}^{L_v(n,k_1, ..., k_{v-1})}
\sum_{i=1}^lF_i(n,n_1,\dots,n_r,k_1,\dots,k_v)
\end{equation}
where the summand is given by hypergeometric terms (e.g., in terms of the Gamma-function) and harmonic sums that arise in the numerators. Then the essential problem is the simplification of these sums to special functions, like, e.g., harmonic sums~\eqref{Equ:HarmonicSums},  generalized harmonic sums~\cite{Moch:02,ABS:13}
\begin{equation}\label{Equ:SSumDef}
	S_{c_1,\ldots ,c_r}(x_1,\ldots ,x_r;k)= \sum_{i_1=1}^k\frac{x_1^{i_1}}{i_1^{c_1}}
\sum_{i_2=1}^{i_1}\frac{x_2^{i_2}}{i_2^{c_2}}\dots\sum_{i_r=1}^{i_{r-1}}\frac{x_r^{i_r}}
{i_r^{c_r}}
\end{equation}
with $c_i\in\NN\setminus\{0\}$ and $x_i\in\set R\setminus\{0\}$, or cyclotomic harmonic sums~\cite{ABS:11}
\begin{align}\label{Equ:CyclotomicSums}
S_{(a_1,b_1,c_1), ...,(a_r,b_r,c_r)}(x_1, ...,x_r; k)
&= \sum_{i_1 = 1}^{k} \frac{x_1^{i_1}}{(a_1 i_1 + b_1)^{c_1}}\sum_{i_2 = 1}^{i_1} \frac{x_2^{i_2}}{(a_2 i_2 + b_2)^{c_2}}\dots \sum_{i_r = 1}^{i_{r-1}} \frac{x_r^{i_r}}{(a_r i_r + b_r)^{c_r}}
\end{align}
where the $a_i, c_i$ are positive integers, the $b_i$ are non-negative integers with $a_i>b_i$, and $x_i\in\set R\setminus\{0\}$. More generally, also binomial sums might arise~\cite{DIS:13}; for a detailed survey see~\cite{AB:2013}.

\noindent For special cases (i.e., if the Gamma-functions in~\eqref{Eq:GenericMultiSum} arise in certain form), this simplification can be carried out with efficient methods; see, e.g.,~\cite{Vermaseren:99,Moch:02,Weinzierl:04}. For more general classes the algorithms and packages presented in this article  have been heavily used~\cite{ACAT:07,Schneider:08e,ACAT:08,BKKS:09,LLPhys:10,DIS:11,ABKSW:11,LLPhys:12,DIS:12,ABHKSW:12,BHKS:13,DIS:13}.

The backbone of all these calculations are the summation tools of \SigmaP\ based on difference fields. In Section~\ref{Sec:Sigma} we will get a glimpse of how the difference field machinery works and elaborate the crucial summation techniques based on this approach.\\ 
In order to apply these tools systematically, the package \texttt{EvaluateMultiSums}~\cite{Schneider:10d,Schneider:13a} has been developed.  In this way, one obtains a completely automatic method to simplify multi-sums of the form~\eqref{Equ:MultiSumsOverIndefNested} in terms of indefinite nested product-sums expressions covering as special cases the sums~\eqref{Equ:HarmonicSums}, \eqref{Equ:SSumDef} and~\eqref{Equ:CyclotomicSums}. The underlying ideas of this package are presented in Section~\ref{Sec:EMS}.\\
During our calculations, e.g., in~\cite{ABKSW:11,BHKS:13,DIS:13}
 we were faced with expressions with several thousand multi-sums. In order to assist this mass-production the package \texttt{SumProduction}~\cite{Schneider:12a} has been developed. In particular, it contains a multi-sum simplifier (again based on the package Sigma) that crunches such large expressions to few master sums and the \texttt{EvaluateMultiSums} package is only applied to those sums. In addition, a useful framework is provided to apply all the calculations in parallel on distributed machines. This latter package is presented in Section~\ref{Sec:SumProduction}.
A conclusion of the symbolic summation approach in difference fields and the produced packages is given in Section~\ref{Sec:Conclusion}.

\section{Symbolic summation in difference fields: the \SigmaP\ package}\label{Sec:Sigma}

An important milestone of symbolic summation is Gosper's telescoping algorithm~\cite{Gosper:78}:
given a hypergeometric expression\footnote{A sequence $f(k)$ (resp.\ an expression that evaluates to a sequence) is called hypergeometric if there is a rational function $r(x)$ and a $\lambda\in\NN$ such that $r(k)=\tfrac{f(k+1)}{f(k)}$ for all $k\in\NN$ with $k\geq\lambda$.} $f(k)$, it finds --in case of existence-- a hypergeometric expression $g(k)$ such that the telescoping equation
\begin{equation}\label{Equ:Tele}
f(k)=g(k+1)-g(k)
\end{equation}
holds. Then given $g(k)$, one can sum~\eqref{Equ:Tele} over $k$ and obtains, e.g., the identity
\begin{equation}\label{Equ:IndefSum}
\sum_{k=1}^af(k)=g(a+1)-g(1).
\end{equation}
Moreover, the breakthrough concerning applications was lead by Zeilberger's extension of Gosper's algorithm to creative telescoping~\cite{Zeilberger:91} in the framework of his holonomic system approach~\cite{Zeilberger:90a}: it enables one to derive recurrence relations for definite hypergeometric sums. In particular, solving such recurrences in terms of hypergeometric expressions~\cite{Petkov:92} gave rise to the following toolbox: given a definite proper hypergeometric sum, one can decide algorithmically if it can be simplified in terms of a linear combination of hypergeometric expressions. For details on this machinery we refer to the pioneering book~\cite{AequalB}; for a most recent point of view see~\cite{KP:11}. 
In the last decades many further improvements and generalizations have been accomplished, like, e.g., for holonomic sequences~\cite{Chyzak:00,Koutschan:13}, for non-holonomic sequences like the Stirling numbers~\cite{Kauers:07} or for expressions represented in terms of difference fields.
The latter approach started with Karr's telescoping algorithm in \pisiSE-fields~\cite{Karr:81}. There one can treat indefinite nested product-sums in a very elegant way; for details see, e.g.,\cite{Schneider:13a}.

\medskip

\noindent\textit{Definition.} Let $f(k)$ be an expression that evaluates at non-negative integers (from a certain point on) to elements of a field $\KK$ containing as subfield the rational numbers $\QQ$. Then $f(k)$ is called indefinite nested product-sum expression w.r.t.\ $k$ (over $\KK$) if it is composed by elements from the rational function field $\KK(k)$, the four operations ($+,-,\cdot,/$), and indefinite
sums and products of the type  
$\sum_{i=l}^kh(i)$ or $\prod_{i=l}^kh(i)$ where $l\in\NN$ and where $h(i)$ is an indefinite nested product-sum expression w.r.t.\ $i$ over $\KK$ which is free of $k$.

\medskip
\noindent Typical examples are the sums given in~\eqref{Equ:HarmonicSums}, \eqref{Equ:SSumDef} and~\eqref{Equ:CyclotomicSums}.

\subsection{The basic mechanism in difference fields}

The basic ideas in the setting of difference fields are presented. This part can be omitted by those readers who are mainly interested in the application of the summation tools.

Consider the following indefinite summation problem: simplify 
$\sum_{k=1}^{n}f(k)$ with $f(k)=k\,S_1(k)$
where $S_1(k)=\sum_{i=1}^k\frac{1}i$ denotes the $k$-th harmonic numbers. To accomplish this task, we hunt for a solution $g(k)$ in terms of $S_1(k)$ such that~\eqref{Equ:Tele} holds. In terms of the shift operator $\myShift_k$ w.r.t.\ $k$ equation~\eqref{Equ:Tele} reads as follows:
\begin{equation}\label{Equ:TeleOp}
f(k)=\myShift_k g(k) - g(k).
\end{equation}
First, the summation objects will be represented step by step in a field $\FF$,  
and along with that the shift operator $\myShift_k$ is rephrased by a field automorphism $\fct{\sigma}{\FF}{\FF}$. 
\begin{myEnumerate}
\item[i)] We start with the rational numbers $\QQ$ and define the (only possible) field automorphism $\fct{\sigma}{\QQ}{\QQ}$ with $\sigma(q)=q$ for all $q\in\QQ$. 
\item[ii)] Next, we need to model $k$ with the shift behavior $\myShift_k k=k+1$: Since all elements in $\QQ$ are constant (i.e., $\sigma(q)=q$ for all $q\in\QQ$), we adjoin a variable $t_1$ to $\QQ$ and extend the automorphism to $\fct{\sigma}{\QQ(t_1)}{\QQ(t_1)}$ with $\sigma(t_1)=t_1+1$. In other words, for $f\in\QQ(t_1)$ the element $\sigma(f)$ is obtained by replacing any occurrence of $t_1$ by $t_1+1$.
\item[iii)] Finally, we represent $S_1(k)$ with the shift behavior $\myShift_k S_1(k)=S_1(k)+\frac{1}{k+1}$: First, we try to find such an element in $\QQ(t_1)$, i.e., we look for a $\gamma\in\QQ(t_1)$ such that $\sigma(\gamma)=\gamma+\frac{1}{t_1+1}$ or equivalently 
$\sigma(\gamma)-\gamma=\frac{1}{t_1+1}$
holds. Using our telescoping algorithms from~\cite{Schneider:13b} (or, e.g., Gosper's algorithm) proves that such an element $\gamma$ does not exist. Therefore we
adjoin the variable $t_2$ to $\QQ(t_1)$ and extend the automorphism $\fct{\sigma}{\QQ(t_1)(t_2)}{\QQ(t_1)(t_2)}$ subject to $\sigma(t_2)=t_2+\frac{1}{t_1+1}$. In other words, for $f\in\QQ(t_1)(t_2)$ the element $\sigma(f)$ can be calculated by replacing any occurrences of $t_1$ by $t_1+1$ and of $t_2$ by $t_2+\frac{1}{t_1+1}$. 
\end{myEnumerate}
In short, we have constructed a difference field $\dfield{\FF}{\sigma}$ with the rational function field $\FF=\QQ(t_1)(t_2)$ together with a field automorphism $\sigma$.
There $k$ and $S_1(k)$ are represented by $t_1$ and $t_2$, respectively, and the shift operator $\myShift_k$ with $\myShift_k k=k+1$ and $\myShift_k S_1(k)=S_1(k)+\frac{1}{k+1}$ is reflected by $\sigma$. 
In this setting, $f(k)$ is given by $\phi=t_1^2\,t_2$ and one seeks $\gamma\in\QQ(t_1)(t_2)$ such that
$\phi=\sigma(\gamma)-\gamma.$
With our algorithm we calculate $\gamma=\frac{1}{4} (t_1-1) t_1 \big(2 t_2-1\big)$ which delivers the solution $g(k)=\frac{1}{4} (k-1)k\big(2 S_1(k)-1\big)$ for~\eqref{Equ:TeleOp} and thus for~\eqref{Equ:Tele}. As a consequence we get~\eqref{Equ:IndefSum} which produces  
the simplification
\begin{equation}\label{Equ:SimpleSId}
\sum_{k=1}^{n}kS_1(k)=\frac{1}{4} \big(2n(n+1) S_1(n)-(n-1)n\big).
\end{equation}

Summarizing, we applied the following strategy; for more details we refer to~\cite{Schneider:13a}.

\begin{myEnumerate}
\item Represent the involved indefinite nested product-sum expressions, whose evaluation leads to elements from a field $\KK$, in a difference field $\dfield{\FF}{\sigma}$. Here $\FF=\KK(t_1)\dots(t_e)$ is a rational function field where the generators $t_i$ represent the sums and products. Moreover, the shift behavior of the objects is described by a field automorphism $\fct{\sigma}{\FF}{\FF}$ where for $1\leq i\leq e$ either the sum relation  $\sigma(t_i)=t_i+a_i$ or the product relation $\sigma(t_i)=a_i\,t_i$ with $0\neq a_i\in\FF_{i-1}:=\KK(t_1)\dots(t_{i-1})$ hold. As indicated in the example above, it is crucial that a new variable $t_i$ with $\sigma(t_i)=t_i+a_i$ (similar for products) is only adjoined to $\FF_{i-1}$ if there is no $\gamma\in\FF_{i-1}$ with $\sigma(\gamma)=\gamma+a_i$. 
Exactly this problem is solvable~\cite{Karr:81}; for improved algorithms see~\cite{Schneider:08c,Schneider:13b} and references therein. 
\item Solve the underlying summation problem (e.g., again telescoping, but also parameterized telescoping and recurrence solving given below) in this setting. 
\item Reformulate the solution to an expression in terms of indefinite nested product-sum expressions that yields a solution of the given summation problem.
\end{myEnumerate}

\noindent Exactly this construction produces difference fields $\dfield{\KK(t_1)\dots(t_e)}{\sigma}$ whose constants remain unchanged;
see~\cite{Karr:81,Karr:85,Schneider:01}. Difference fields with this property are also called \pisiSE-fields. Formally, they can be defined as follows.

\medskip

\textit{Definition.} Let $\FF$ be a field with characteristic
$0$ (i.e., the rational numbers are contained as sub-field) and let $\sigma$ be a field automorphism of $\FF$. The constant field of
$\FF$ is defined by $\KK=\{f\in\FF|\sigma(f)=f\}$. A difference
field $(\FF,\sigma)$ with constant field $\KK$ is called a
\pisiSE-field if $\FF=\KK(t_1)\dots(t_e)$ where for all $1\leq i\leq
e$ each $\FF_i=\KK(t_1)\dots(t_i)$ is a transcendental field
extension of $\FF_{i-1}=\KK(t_1)\dots(t_{i-1})$ (we set $\FF_0=\KK$)
and $\sigma$ has the property that $\sigma(t_i)=a\,t_i$ or
$\sigma(t_i)=t_i+a$ for some $a\in\FF_{i-1}$.

\medskip

In conclusion, all the summation paradigms presented in the next subsections rely on this mechanism: Reformulate the given problem in a \pisiSE-field (or ring), solve it there and formulate the result back such that it is a solution of the input problem.
In the following we will give an overview of \SigmaP's symbolic summation toolbox that is based on the difference field (resp.\ ring) approach introduced above.

\subsection{Simplification of indefinite nested product-sum expressions}
\label{Sec:IndefSum}

Whenever \SigmaP\ deals with indefinite nested product-sum expressions, in particular when it outputs such expressions, the following problem is solved implicitly. 

\bigskip

\noindent\MyFrame{\noindent\textbf{Problem EAR: Elimination of algebraic relations.}
\textit{Given} an indefinite nested product-sum expression $f(k)$.
\textit{Find} an indefinite nested product-sum expression $F(k)$ and $\lambda\in\NN$ such that $f(k)=F(k)$ for all $k\geq\lambda$ and such that the occurring sums are algebraically independent.
}

\bigskip

\noindent As worked out above, the following mechanism is applied: $f(k)$ is rephrased in a suitable \pisiSE-field $\dfield{\KK(t_1)\dots(t_e)}{\sigma}$ (or ring); this is accomplished by solving iteratively the telescoping problem. Here the sums (similarly the products) are represented by the variables $t_i$. Finally, reinterpreting the $t_i$ as sums produces an alternative expression $F(k)$ of $f(k)$ where the occurring sums are algebraically independent; the solution to this problem relies on results of~\cite{Schneider:10c} and is worked out in detail in~\cite{Schneider:13a}.
We remark that that the found relations for harmonic sums, cyclotomic sums and generalized harmonic sums coincide with the derived relations~\cite{Bluemlein:04,ABS:11,ABS:13,ABS:13b} that are obtained by using the underlying quasi-shuffle algebras~\cite{Vermaseren:99,Hoffman:00,Moch:02}.

\medskip

E.g., after loading the \SigmaP\ package into Mathematica: 

\begin{Cmma}
\CIn << Sigma.m\\\vspace*{-0.1cm}
\CPrint Sigma - A\; summation\; package\; by\; Carsten\;\; Schneider \copyright\ RISC\\
\end{Cmma}

\noindent the simplification in~\eqref{Equ:SimpleSId} can be accomplished as follows\footnote{Here $S[c_1,\dots,c_r,k]$ denotes the harmonic sum~\eqref{Equ:HarmonicSums}.}:

\begin{Cmma}
\CIn SigmaReduce[\sum_{k=1}^{n}k\,S[1,k],n]\\
\COut \frac{1}{4} \big(2n(n+1) S[1,n]-(n-1)n\big)\\
\end{Cmma}

\noindent  Here the sum $\sum_{k=1}^{n}k\,S_1(k)$ has been reduced in terms of the objects $k$ and $S_1(k)$. In particular, by difference field theory it follows that the sequence given by $S_1(k)$ is transcendental (resp.\ algebraically independent) over the rational sequences, i.e., the sequences  that one obtains by evaluating the elements of $\QQ(k)$. 

\medskip

In particular, using improved telescoping algorithms, the underlying \pisiSE-field can be constructed in such a way that the sums are given with certain optimality criteria.

\medskip

\noindent\MyFrame{\noindent\textbf{Refined telescoping.} 
\textit{Given} an indefinite nested product-sum expression $f(k)$.\\ 
\textit{Find} an indefinite nested product-sum expression $g(k)$ such that~\eqref{Equ:Tele} holds and such that $g(k)$ is as ``simple'' as possible.}

\medskip

\noindent Then summing the found equation~\eqref{Equ:Tele} over $k$ yields, e.g., the identity~\eqref{Equ:IndefSum} where also the right hand side is as simple as possible. Subsequently, we present the main features of \SigmaP; some of these simplifications are carried out by default, some must be activated explicitly (see below).


\subsubsection{Sum representations with optimal nesting depth} The found expression $F(k)$ can be given with minimal nesting depth and for any occurring sum in $F(k)$ there is no other indefinite nested sum representation with lower nesting depth; for details see~\cite{Schneider:05f,Schneider:08c,Schneider:10b}. A typical example is as follows.

\begin{Cmma}\CMLabel{MMA:Indef}
\CIn SigmaReduce[2^n\sum_{i=1}^n \Mfrac{1}{i}\sum_{j=1}^i \Mfrac1{2^{j}} \sum_{k=1}^j \Mfrac{2^k}{k}\sum_{l=1}^k \Mfrac1{2^{l}},n]\vspace*{-0.3cm}\\
\COut 2^n \big(\big(\sum_{i=1}^n \frac{1}{i}\big)^2+\big(\sum_{i=1}^n \Mfrac1{2^{i}i}\big) \sum_{i=1}^n \frac{1}{i}+\sum_{i=1}^n \frac{1}{i^2}+\sum_{i=1}^n \Mfrac1{2^{i}i^2}-3 \sum_{i=1}^n \Mfrac{\sum_{j=1}^i \Mfrac1{2^{j}j}}{i}-\sum_{i=1}^n \Mfrac{\Mfrac1{2^{i}} \sum_{j=1}^i \frac{2^j}{j}}{i}\big)\\
\end{Cmma}

\smallskip

\noindent Here the product $2^n=\prod_{i=1}^n2$ and the arising sums are algebraically independent over the rational sequences, and the sums have minimal nesting depth. In particular, the expression in~\CmyOut{\ref{MMA:Indef}} cannot be written in terms of indefinite nested sums with lower nesting depth. Similarly, the following sum expression can be reduced to its optimal nesting depth.

\begin{Cmma}
\CIn SigmaReduce[\sum_{k=0}^a \Big(\sum_{j=0}^k 
\frac{x^j}{\binom{n}{j}}\Big)^2]\\
\COut \frac{-2x^{a+1}(a+1)}{(x+1) \binom{n}{a}} \sum_{i_1=0}^a
\frac{x^{i_1}}{\binom{n}{i_1}}+\frac{(n+x+1)}{x+1}\sum_{i_1=0}^a
\frac{x^{2 i_1}}{\binom{n}{i_1}{}^2}+\frac{(x-1)}{x+1}\sum_{i_1=0}^a
\frac{x^{2 i_1} i_1}{\binom{n}{i_1}{}^2}+\frac{(a x+a-n+2x)}{x+1}\Big(\sum_{i_1=0}^a \frac{x^{i_1}}{\binom{n}{i_1}}\Big){}^2\\
\end{Cmma}

\noindent We emphasize that this depth-optimal representation leads to efficient algorithms for telescoping, creative telescoping and recurrence solving; for details we refer to~\cite{Schneider:08c,Schneider:10a}. Thus by efficiency reasons this feature of \SigmaP\ is activated by default.

\subsubsection{Reducing the number of objects and the degrees in the summand} The depth-optimal representation can be refined further as follows: \textit{find} an alternative sum representation such that for the outermost summands the number of occurring objects is as small as possible. This problem was originally solved in~\cite{Schneider:04a}. For a more efficient and simplified algorithm see~\cite{Schneider:13b}.

\noindent E.g., in the following example we can eliminate $S_1(k)$ from the summand:

\begin{Cmma}
\CIn SigmaReduce[\sum_{k=0}^a (-1)^k S[1,k]^2 \binom{n}{k},a,SimplifyByExt\to DepthNumber]\\\vspace*{-0.1cm}
\COut -(a-n) \big(n^2 S[1,a]^2+2\,n\,S[1,a]+2\big) \frac{(-1)^a \binom{n}{a}}{n^3}-\frac{2}{n^2}-\frac{1}{n}\sum_{i_1=1}^a \frac{(-1)^{i_1}}{i_1}\binom{n}{i_1}\\
\end{Cmma}

\noindent Furthermore, one can calculate representations such that the degrees (w.r.t.\ the top extension of a \pisiSE-field) in the numerators and denominators of the summands are minimal~\cite{Schneider:07d}:

\begin{Cmma}
\CIn SigmaReduce[\sum_{k=0}^a (-1)^k S[1,k]{}^3 \binom{n}{k}]\\
\COut-\frac{3}{n^2} \sum_{i_1=1}^a \frac{(-1)^{i_1} \binom{n}{i_1}}{i_1}-\frac{3}{n} \sum_{i_1=1}^a \frac{(-1)^{i_1} \binom{n}{i_1} S[1,i_1]}{i_1}+\frac{1}{n}\sum_{i_1=1}^a \frac{(-1)^{i_1} \binom{n}{i_1}}{i_1^2}+(-1)^a \binom{n}{a} \big(\frac{6 (n-a)}{n^4}+\frac{6 (n-a) S[1,a]}{n^3}-\frac{3 (a-n) S[1,a]^2}{n^2}+\frac{(n-a) S[1,a]^3}{n}\big)-\frac{6}{n^3}\\
\end{Cmma}

\noindent For algorithms dealing with the product case we point to~\cite{Schneider:05c,Petkov:10}.

\subsubsection{Minimal degrees w.r.t.\ the summation index} By default \SigmaP\ outputs sums such that the denominators have minimal degrees w.r.t.\ the summation index (i.e., if possible, the denominator w.r.t.\ the summation index is linear). This feature is of particular importance to rewrite sums in terms of harmonic sums and their generalized versions. A typical example is

\begin{Cmma}
\CIn SigmaReduce[\sum_{k=1}^a \Big(\frac{-2+k}{10 (1+k^2)}+\frac{(1-4 k-2 k^2)S[1,k]}{10 (1+k^2) (2+2 k+k^2)}+\frac{(1-4 k-2 k^2)S[3,k]}{5 (1+k^2) (2+2 k+k^2)}\Big),a]\\\vspace*{-0.05cm}
\COut \frac{a^2+4 a+5}{10 (a^2+2 a+2)}S[1,a]-\frac{(a-1)(a+1)}{5(a^2+2a+2)}S[3,a]-\tfrac{2}{5}\sum_{k=1}^a\frac{1}{k^2}\\
\end{Cmma}

\subsection{The summation paradigms for definite summation}
\label{Sec:DefSummation} 
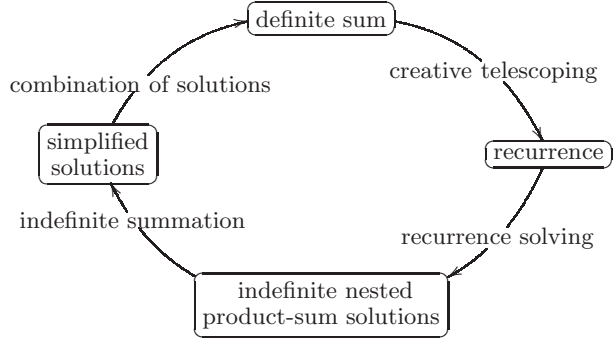
\begin{figure}
  \caption{\SigmaP's summation spiral\label{Fig:Spiral}; see~\cite{Schneider:04c}.}
  \centering
\footnotesize
$$\xymatrix@R=1.2cm@C=0.5cm{
&*+[F-:<3pt>]{\txt{definite sum}} \ar @/^1.5pc/[rd]|>>>>>>>>>>{\txt{creative telescoping}}&\\
*+[F-:<3pt>]{\txt{simplified\\ solutions}}\ar @/^1.5pc/[ru]|>>>>>>>>>>>{\txt{combination of
solutions}} &&*+[F-:<3pt>]{\txt{recurrence}}\ar @/^1.5pc/[ld]|<<<<<<<<<<{\txt{recurrence solving}}\\
&*+[F-:<3pt>]{\txt{ 
indefinite nested\\ 
product-sum solutions}}\ar @/^1.5pc/[lu]|<<<<<<<<<{\txt{indefinite
summation}}& }$$
\normalsize
\end{figure}

Definite sums over indefinite nested product sums\footnote{Definite means that the upper bound is $\infty$ or consists of parameters that occur also inside of the sum.}, like
\begin{equation}\label{Equ:DefSum}
A(n)=\sum_{k=0}^n \binom{n}{k}S_1(k)^2=\sum_{k=0}^n \Big(\prod_{i=1}^k\frac{n+1-i}i\Big)\Big(\sum_{i=1}^k\frac{1}i\Big)^2,
\end{equation}

\noindent can be handled by the following summation paradigms (see Fig.~\ref{Fig:Spiral}). 

\subsubsection{Finding recurrences by parameterized (creative) telescoping}

First, there is the following tool in the setting of \pisiSE-fields to obtain recurrences; for the most recent summary see~\cite{Schneider:13b}.

\bigskip

\noindent\MyFrame{\noindent\textbf{Problem PT: Parameterized Telescoping.}
\textit{Given} indefinite nested product-sum expressions $f_0(k),\dots,f_{\delta}(k)$.
\textit{Find} constants $a_0,\dots,a_{\delta}$, not all 0 and all free of $k$, and \textit{find} an indefinite nested product-sum expression $g(k)$ being not more complicated than the $f_i(k)$ such that
\begin{equation}\label{Equ:Crea}
a_0\,f_0(k)+a_1\,f_1(k)+\dots+a_{\delta}\,f_{\delta}(k)=g(k+1)-g(k).
\end{equation}
}

\bigskip

\noindent For simplicity suppose that the found relation~\eqref{Equ:Crea} holds for $0\leq k\leq a$. Then by telescoping one gets, e.g., the sum relation
\begin{equation}\label{Equ:ParaRec}
a_0\,\sum_{k=0}^af_0(k)+a_1\,\sum_{k=0}^af_1(k)+\dots+a_{\delta}\,\sum_{k=0}^af_{\delta}(k)=g(a+1)-g(0)
\end{equation}
where the right hand side is simpler than the sums of the left hand side.

\noindent Specializing to $f_i(k):=f(n+i,k)$ for a bivariate expression yields the creative telescoping paradigm. Here one loops over\footnote{Note that the special case $\delta=0$ is telescoping.} $\delta=0,1,2,\dots$ and tries to solve the corresponding parameterized telescoping problem. If the method stops, then we can deduce~\eqref{Equ:ParaRec}, i.e., we obtain the recurrence
\begin{equation*}
a_0(n)\,A'(n)+a_1(n)\,A'(n+1)+\dots+a_{\delta}(n)\,A'(n+\delta)=g(a+1)-g(0)
\end{equation*}
for the sum $A'(n)=\sum_{k=0}^af(n,k)$. To this end, specializing $a=n$ and taking care of extra terms yield a recurrence for the sum $A(n)=\sum_{k=0}^nf(n,k)$ of the form
\begin{equation}\label{Equ:Rec}
a_0(n)A(n)+a_1(n)A(n+1)+\dots+a_{\delta}(n)A(n+\delta)=h(n).
\end{equation}

E.g., take $f_i(k)=\binom{n+i}{k}S_1(k)^2=\prod_{j=1}^i\frac{n+j}{n-k+j}\binom{n}{k}S_1(k)^2$. Then we find a parameterized telescoping solution~\eqref{Equ:Crea} for $\delta=4$. 
Performing the steps above, we finally get a recurrence of order 4 for our sum~\eqref{Equ:DefSum}. All these steps can be carried out for $A(n)=\texttt{SUM[n]}$ as follows:

\begin{Cmma}
\CIn mySum=\sum_{k=1}^n \binom{n}{k} S[1,k]^2;\\
\end{Cmma}
\begin{Cmma}\CMLabel{MMA:Rec}
\CIn rec=GenerateRecurrence[mySum,n][[1]]\\
\COut 8 (n+1) (n+3) \text{SUM}[n]-4 \big(5 n^2+25 n+29\big) \text{SUM}[n+1]+2 (3 n+8) (3 n+10) \text{SUM}[n+2]\newline
-\big(7 n^2+49 n+86\big) \text{SUM}[n+3]+(n+4)^2 \text{SUM}[n+4]=1\\
\end{Cmma}


\medskip

We remark that the refined telescoping algorithms from Subsection~\ref{Sec:IndefSum} provide also refined tools for parameterized/creative telescoping~\cite{Schneider:08c,Schneider:13b}. E.g., for 
our sum~\eqref{Equ:DefSum} we can calculate a recurrence of order 2 instead of order 4 by using extra sum extensions:

\begin{Cmma}
\CIn GenerateRecurrence[mySum,n,SimlifyByExt\to DepthNumber][[1]]\\
\COut 4 (n+1) \text{SUM}[n]-2 (2 n+3) \text{SUM}[n+1]+(n*2) \text{SUM}[n+2]==\frac{(3 n+4)}{n+2}\sum_{i_1=0}^n \frac{\binom{n}{i_1}}{1+n-i_1}+\sum_{i_1=1}^n \frac{\binom{n}{i_1}}{i_1}+\frac{1}{n+2}\\
\end{Cmma}

\noindent Note that the found sums on the right hand side are definite. Simplifying these sums (by using just the methods that we describe here), we end up at the recurrence
$$4 (n+1) A(n)-2 (2 n+3) A(n+1)+(n+2) A(n+2)
=S_1(2;n)-S_1(n)+\tfrac{2^{n+1} (3 n+4)-(2 n+3)}{(n+1) (n+2)}.
$$
Usually such shorter recurrences are easier to solve. In order to demonstrate the summation tools below, we will continue with the recurrence given in~\CmyOut{\ref{MMA:Rec}}.

\subsubsection{Solving recurrences} Next, we can apply the following recurrence solver~\cite{ABPS:13} which is based on~\cite{Bron:00,Schneider:05a} and generalizes ideas of~\cite{Petkov:92,Abramov:94}.

\medskip

\noindent\MyFrame{\noindent\textbf{Problem RS: Recurrence Solving.}\\
\textit{Given} a recurrence of the form~\eqref{Equ:Rec}
where the coefficients $a_i(n)$ and $h(n)$ are given in terms of indefinite nested product-sum expressions. \textit{Find} all solutions that are expressible in terms of indefinite nested product sum expressions.
}

\medskip

\noindent This solver finds, if possible, a particular solution of~\eqref{Equ:Rec} in terms of indefinite nested product-sum expressions; and it finds a linear independent set of expressions in terms of indefinite nested product-sums with the following property: their linear combinations produce all solutions of the homogeneous version of~\eqref{Equ:Rec} that can be expressed in terms of indefinite nested product-sum expressions. The sequences generated by these solutions are called d'Alembertian solutions, a subclass of Liouvillian solutions~\cite{Singer:99}. For details dealing with the rational case see~\cite{Petkov:2013}.

E.g., by executing the following command with the recurrence~\texttt{rec}=\CmyOut{\ref{MMA:Rec}} 

\begin{Cmma}
\CIn recSol=SolveRecurrence[rec,SUM[n],IndefiniteSummation\to False]\vspace*{-0.3cm}\\
\COut \{ \{0, 2^n\},
 \{0, \displaystyle2^n \sum_{i=1}^n \frac{1}{i}\},
 \{0 , 2^{n} \displaystyle\sum_{i=1}^n \Mfrac{\sum_{j=1}^i \Mfrac{1}{2^{j}}}{i}\},
 \{0 , 2^n \displaystyle\sum_{i=1}^n \Mfrac{\sum_{j=1}^i \Mfrac1{2^{j}}\sum_{k=1}^j \frac{2^k}{k}}{i}\},
 \{1 , 2^n\displaystyle \sum_{i=1}^n \Mfrac{\sum_{j=1}^i \Mfrac1{2^{j}} \sum_{k=1}^j \Mfrac{2^k \sum_{l=1}^k \Mfrac1{2^{l}}}{k}}{i}\}\}\\
\end{Cmma}

\noindent we obtain the general solution
\small
$$2^n\Bigg[c_1+c_2\,\sum_{i=1}^n \frac{1}{i}+c_3\,\displaystyle\sum_{i=1}^n \frac1i\sum_{j=1}^i \frac{1}{2^{j}}+c_4\displaystyle\sum_{i=1}^n \Mfrac{1}{i}\sum_{j=1}^i \frac1{2^{j}}\sum_{k=1}^j \frac{2^k}{k}
+\displaystyle \sum_{i=1}^n \frac{1}{i}\sum_{j=1}^i \frac1{2^{j}}\sum_{k=1}^j \frac{2^k}{k} \sum_{l=1}^k \frac1{2^{l}}\Bigg],\;\; c_1,c_2,c_3,c_4\in\QQ.$$
\normalsize
\noindent By construction the solutions are highly nested: e.g., the depth of the particular solution equals usually the recurrence order plus the nesting depth of the inhomogeneous part of the recurrence. It is therefore a crucial (and often the most challenging) task to simplify these solutions further by solving Problem~EAR and 
applying the refined telescoping algorithms from Subsection~\ref{Sec:IndefSum}.
For the simplification of the particular solution see~\CmyOut{\ref{MMA:Indef}} from above. With \SigmaP\ the recurrence \texttt{rec} given in~\CmyOut{\ref{MMA:Rec}} is solved (see Problem RS) and the found solutions are simplified (with the default options) by the following function call:

\begin{Cmma}
\CIn recSol=SolveRecurrence[rec,SUM[n]]\\
\COut \{\{ 0 , 2^n\},
 \{0 , 2^n \sum_{i=1}^n \frac{1}{i}\},
 \{0 , 2^n \big(2 \sum_{i=1}^n \frac{1}{i}-2 \sum_{i=1}^n \Mfrac1{2^{i}i}\big)\},
 \{0 , 2^n \big(\big(\sum_{i=1}^n \frac{1}{i}\big)^2+\sum_{i=1}^n \frac{1}{i^2}-\sum_{i=1}^n \Mfrac1{2^{i}i} \sum_{j=1}^i \Mfrac{2^j}{j}\big)\},\vspace*{-0.1cm}\newline
 \{1 , 2^n \big(\big(\sum_{i=1}^n \frac{1}{i}\big)^2+\sum_{i=1}^n \frac{1}{i^2}+\big(\sum_{i=1}^n\frac{1}{i}\big) \sum_{i=1}^n\Mfrac1{2^{i}i} +\sum_{i=1}^n \Mfrac1{2^{i}i^2}-\sum_{i=1}^n\Mfrac1{2^{i}i} \sum_{j=1}^i \frac{2^j}{j}-3 \sum_{i=1}^n \Mfrac1i\sum_{j=1}^i \Mfrac1{2^{j}j}\big)\}\}\\
\end{Cmma}

\noindent\textit{Example.} In~\cite{ACAT:08,BKKS:09} we computed a large amount of initial values (up to 3500) of the massless
Wilson coefficients to 3-loop order for individual color coefficients by taking the result given in~\cite{Moch:04}. Next, we guessed recurrences of minimal order for these coefficients by using Kauer's package \texttt{Guess}~\cite{Guessing:09}. Then, e.g., the largest recurrence of order 35 could be solved completely in about 1 day. This yielded 35 linearly independent solutions in terms of sums up to nesting depth 34. To this end, their simplifications in terms of harmonic sums took about 5 days.


\subsubsection{Combining the solutions}
Finally, we take the linear combination of the homogeneous solutions (the entries with a $0$) plus the particular solution (the entry with a $1$) such that it agrees with $A(n)$ for $n=1,2,3,4$. This combination can be calculated by calling the function

\begin{Cmma}\CMLabel{MMA:LinComb} 
\CIn sol=FindLinearCombination[recSol,mySum,n,4]\\
\COut 2^n \big(\big(\sum_{i=1}^n \frac{1}{i}\big)^2+\sum_{i=1}^n \frac{1}{i^2}+\big(\sum_{i=1}^n\frac{1}{i}\big) \sum_{i=1}^n\Mfrac1{2^{i}i} +\sum_{i=1}^n \Mfrac1{2^{i}i^2}-\sum_{i=1}^n\Mfrac1{2^{i}i} \sum_{j=1}^i \frac{2^j}{j}-3 \sum_{i=1}^n \Mfrac1i\sum_{j=1}^i \Mfrac1{2^{j}j}\big)\\
\end{Cmma}

\noindent Since the sum~\eqref{Equ:DefSum} and the derived expression in~\CmyOut{\ref{MMA:LinComb}} agree for $n=1,2,3,4$ and since both are solutions of the recurrence~\CmyOut{\ref{MMA:Rec}}, they evaluate to the same sequence for all $n\in\NN$.

In order to rewrite the found expression~\CmyOut{\ref{MMA:LinComb}} in terms of harmonic sums and their generalized versions, we load in the \texttt{HarmonicSums} package and execute the following function

\begin{Cmma}
\CIn << HarmonicSums.m \\
\CPrint HarmonicSums\; by\; Jakob\;\; Ablinger -- \copyright\ RISC\\
\end{Cmma}

\vspace*{-0.6cm}

\begin{Cmma}
\CIn TransformToSSums[sol]\\
\COut 2^n \Big(S[1,n]^2+S[2,n]+S[1,n] S[1,\{\tfrac{1}{2}\},n]\big)+S[2,\{\tfrac{1}{2}\},n]-S[1,1,\{\tfrac{1}{2},2\},n]-3 S[1,1,\{1,\tfrac{1}{2}\},n]\Big)\\
\end{Cmma}

\noindent To sum up, using the summation paradigms given in Fig.~\ref{Fig:Spiral}, we computed for the definite sum~\eqref{Equ:DefSum} the closed form in terms of generalized harmonic sums:
\begin{equation}\label{Equ:DefId}
A(n)=2^n \big[S_1(n)^2+S_2(n)+S_1(n) S_1\big(\tfrac{1}{2};n\big)+S_2\big(\tfrac{1}{2};n\big)-S_{1,1}\big(\tfrac{1}{2},2;n\big)-3 S_{1,1}\big(1,\tfrac{1}{2};n\big)\big)].
\end{equation}

\noindent\textit{Remark.} We highlight that all the calculation steps can be verified independently of the way how the (complicated) algorithms work. In this way, we obtain rigorous computer proofs.

\section{Automatic simplification of multiple sums: the \texttt{EvaluateMultiSums} package}\label{Sec:EMS}

Using \SigmaP's summation tools (see Fig.~\ref{Fig:Spiral})  the derivation of the right hand side of~\eqref{Equ:DefId} can be done completely automatically with the package 
\begin{Cmma}
\CIn << EvaluateMultiSums.m \\
\CPrint EvaluateMultiSums\; by\; Carsten\;\; Schneider - \copyright\ RISC\\
\end{Cmma}

\vspace*{-0.3cm}

\noindent Namely, our sum~\eqref{Equ:DefSum} can be simplified to indefinite nested sums by executing
\begin{Cmma}
\CIn EvaluateMultiSum[\sum_{k=1}^nS[1,k]^2 \binom{n}{k},\{\{k,0,n\}\},\{n\},\{0\},\{\infty\}]\\
\COut 2^n \Big(S[1,n]^2+S[2,n]+S[1,n] S[1,\{\tfrac{1}{2}\},n]\big)+S[2,\{\tfrac{1}{2}\},n]-S[1,1,\{\tfrac{1}{2},2\},n]-3 S[1,1,\{1,\tfrac{1}{2}\},n]\Big)\\
\end{Cmma}

\noindent The underlying method applies systematically the tools of \SigmaP\ (taking care of all the different options to find the optimal treatment) until the simplification is accomplished. 

The key point is that this method can be applied iteratively to multiple sums. E.g., in recent calculations we succeeded in deriving the massive 3-loop OMEs $A_{qq,Q}^{(3),\rm NS}$ and 
$A_{qq,Q}^{(3),\rm NS,TR}$ for general values of $N$, in particular we obtained the Wilson coefficient 
$L_{qq,Q}^{(3),\rm NS}$; for further comments we refer to~\cite{DIS:13}. Here one (of many multi-sums) is the hypergeometric quadruple sum 
\begin{equation}\label{Equ:4SumEp}
{\cal I}(\ep,n)=\sum_{m=0}^n\sum_{j=0}^m\sum_{k=0}^{\infty}\sum_{i=0}^k f(\ep,n,m,j,k,i)
\end{equation}
where the summand is given in terms of Gamma functions and binomial coefficients:
\begin{Cmma}
\CIn f=\tfrac{(-1)^{i+m+1} e^{-\frac{3 \ep \gamma }{2}} \binom{k}{i} \binom{m}{j} \binom{N}{m}\Gamma \big(1-\frac{\ep}{2}\big) \Gamma \big(\frac{\ep}{2}\big) \Gamma \big(i-\frac{\ep}{2}\big)}{k! \Gamma \big(-\frac{\ep}{2}+i+1\big) \Gamma \big(-\frac{3 \ep}{2}+i+j+2\big)} \tfrac{\Gamma (-\ep+i+j+1) \Gamma \big(k-\frac{3 \ep}{2}\big) \Gamma \big(\frac{\ep}{2}+j+k+1\big) \Gamma (m+2) \Gamma (-\ep-j+m+1)}{\Gamma \big(\frac{\ep}{2}+m+2\big) \Gamma \big(-\frac{\ep}{2}+k+m+2\big)}\\
\end{Cmma}
\noindent  In this instance, the first 5 coefficients $I_{-3}(n),\dots,I_1(n)$ of its Laurent series expansion~\eqref{Equ:LaurentExp} with $t=-3$ were needed.
For this task (see Section~\ref{Sec:Intro}) we first expand the summand 
\begin{equation}\label{Equ:ConcreteSummand}
f(\ep,n,m,j,k,i)=F_{-3}(n,m,j,k,i)\ep^{-3}+\dots+F_{1}(n,m,j,k,i)\ep^{1}+O(\ep^2).
\end{equation}
Then the coefficients $I_i(n)$ are given by applying the sums to the $F_i(n,m,j,k,i)$.
E.g., we compute $F_{-1}(n,m,j,k,i)$ by using the following function of the package \texttt{EvaluateMultiSums}:

\begin{Cmma}
\CIn F_{-1}=SeriesForProduct[f,\{\ep,-1,-1\}]\\
\COut \frac{(-1)^{i+m+1} 2\binom{k}{i} \binom{m}{j} \binom{n}{m} (i+j)! (k-1)! (j+k)! (m-j)!}{i (i+j+1)! k! (k+m+1)!}\\
\end{Cmma}

\noindent In other words, the main task is to simplify the definite multi-sum

\vspace*{-0.4cm}

\begin{equation}\label{Equ:4FoldEpFree}
\overbrace{\sum_{m=0}^n}^{(d)}\overbrace{\sum_{j=0}^m}^{(c)}\overbrace{\sum_{k=0}^{\infty}}^{(b)}\overbrace{\sum_{i=1}^k}^{(a)} \frac{(-1)^{i+m+1} 2\binom{k}{i} \binom{m}{j} \binom{n}{m} (i+j)! (k-1)! (j+k)! (m-j)!}{i (i+j+1)! k! (k+m+1)!};
\end{equation}

\vspace*{-0.2cm}

\noindent note that the inner sum (a) starts with $i=1$ since there is a pole at $i=0$. Exactly here (as for all the other multi-sums in this context) the presented toolbox can be exploited. The sum (a) is a definite sum over an indefinite nested product-sum. Therefore we can can activate our definite summation technology (Fig.~\ref{Fig:Spiral}): try to transform it to an indefinite nested expression w.r.t.\ $k$ (which is the summation index of the next sum (b)). Namely, we execute
the function\footnote{The free integer parameters $k,j,m,n$ are given explicitly where their lower values are $0,0,0,2$ and their upper values are $\infty,m,n,\infty$, respectively. I.e., $0\leq k\leq\infty$, $0\leq j\leq m$, $0\leq m\leq \infty$, $0\leq n\leq\infty$. In particular, the given order $k,j,m,n$ specifies how the sums should be transformed: if a sum depends on $k$, it should be transformed to indefinite sums w.r.t.\ $k$; if it is free of $k$, but depends on $j$, it should be transformed w.r.t.\ $j$, etc.
}:

\begin{Cmma}
\CIn sumA=EvaluateMultiSum[F_{-1}, \{\{i, 1, k\}\}, \{k, j, m, n\}, \{0, 0, 0, 
  2\}, \{\infty, m, n, \infty\}]\\
\COut -2 \frac{(-1)^m m!}{(k+m+1)!} \big(\big(\frac{1}{(j+1)^2 k}-\frac{S[1,k]}{(j+1) k}\big) \frac{\binom{n}{m} (j+k)!}{j!}-\frac{k! \binom{n}{m}}{(j+1) k (j+k+1)}\big)\\
\end{Cmma}

\noindent Given this form, the next sum (b) fits again to our summation paradigm: it is a definite sum over an indefinite nested product sum-expression w.r.t.\ $k$. Hence with the function call

\begin{Cmma}
\CIn sumB=EvaluateMultiSum[sumA, \{\{k, 0, \infty\}\}, \{j, m, n\}, \{0, 0, 
  2\}, \{m, n, \infty\}]\\
\COut \frac{(-1)^m \binom{n}{m}}{(j+1) (m+1)} \big(\sum_{\text{i}_ 1=1}^j \frac{1}{1+m-\text{i}_ 1}\big)^2-2 \frac{(-1)^m \binom{n}{m}}{(j+1)^2 (m+1)} \sum_{\text{i}_ 1=1}^j \frac{1}{1+m-\text{i}_ 1}+\frac{2 (-1)^m \binom{n}{m}}{(j+1)^3 (m+1)}\newline
+(-1)^m \big(2 \frac{\binom{n}{m} (-m)_j}{(j+1)^2 j!}-\frac{2 \binom{n}{m}}{(j+1) (m+1)}\big) S[2,m]+\frac{(-1)^m \binom{n}{m}}{(j+1) (m+1)} \sum_{\text{i}_ 1=1}^j \frac{1}{\big(1+m-\text{i}_ 1\big)^2}-2 \frac{(-1)^m \binom{n}{m} (-m)_j}{(j+1)^2 j!} \sum_{\text{i}_ 1=1}^j \frac{\text{i}_ 1!}{(-m)_ {\text{i}_ 1} \text{i}_ 1^2}\newline
+2 \frac{(-1)^m \binom{n}{m}}{(j+1) (m+1)} \sum_{\text{i}_ 1=1}^j \frac{\sum_{\text{i}_ 2=1}^{\text{i}_ 1} \frac{1}{1+m-\text{i}_ 2}}{\text{i}_ 1}+(-1)^m \big(\frac{2 \binom{n}{m}}{(j+1) (m+1)}-2 \frac{\binom{n}{m} (-m)_j}{(j+1)^2 j!}\big) \text{z}_ 2\\
\end{Cmma}

\noindent we obtain an indefinite nested product-sum expression w.r.t.\ $j$ (if sums are free of $j$, they are indefinite nested w.r.t. $m$, etc.). Again we are ready to apply our summation toolkit:

\begin{Cmma}
\CIn sumC=EvaluateMultiSum[sumB, \{\{j, 0, m\}\}, \{m, n\}, \{0, 
  2\}, \{n, \infty\}]\\
\COut (-1)^m \big(2 \frac{(-1)^m (-n)_m}{(m+1)^4 m!}+2 \frac{(-1)^m
(-n)_m}{(m+1)^3 m!} S[1,m]-2 -\frac{(-1)^m (-n)_m}{(m+1) m!} S[2,1,m]\big)\\
\end{Cmma}

\noindent This yields an indefinite nested product sum expression w.r.t.\ $m$. To this end, the outermost sum $(d)$ is transformed to an indefinite nested product-sum expression
\begin{Cmma}
\CIn sumD=EvaluateMultiSum[sumC, \{\{m, 0, n\}\}, \{n\}, \{2\}, \{\infty\}]\\
\COut \frac{2 S[2,n]}{(n+1)^2}+\frac{2}{(n+1)^4}\\
\end{Cmma}

\noindent which is nothing else than the simplification of~\eqref{Equ:4FoldEpFree}.
The full power of this machinery comes into action if the function is applied in one stroke by the following function call:

\begin{Cmma}
\CIn EvaluateMultiSum[F_{-1},\{\{i,1,k\},\{k,0,\infty\},\{j,0,m\},\{m,0,N\}\},\{n\},\{2\},\{\infty\}]\\
\COut \frac{2 S[2,n]}{(n+1)^2}+\frac{2}{(n+1)^4}\\
\end{Cmma}

More generally, we can deal with the following problem.

\medskip

\noindent\MyFrame{\noindent\textbf{Problem EMS: EvaluateMultiSum.}
\textit{Given} 

\vspace*{-0.2cm}

\begin{equation}\label{Equ:EMSInputSum}
F(\vect{m})=\sum_{k_1=l_1}^{L_1(\vect{m})} ... \sum_{k_v=l_v}^{L_v(\vect{m},k_1, ..., k_{v-1})}
f(\vect{m},k_1,\dots,k_v)
\end{equation}
\noindent with an indefinite nested product-sum expression $f$ w.r.t.\ $k_v$, integer parameters $\vect{m}=(m_1\dots,m_r)$; $l_i\in\NN$ and $L_i(\dots)$ stands for $\infty$ or a linear combination of the involved parameters with integer coefficients.
\textit{Find} an indefinite nested product-sum expression\footnote{If a sum depends on $m_r$, it should occur only in the outermost bound. If it is free of $m_r$, but depends on $m_{r-1}$, the parameter $m_{r-1}$ should only occur in the outermost bound, etc.} $\bar{F}(\vect{m})$ which evaluates to the same expression as $F(\vect{m},n)$.}

\medskip

\noindent Moreover, if one uses, e.g., the option \texttt{ExpandIn$\to\{\ep,-3,1\}$} also the expansion feature is applied, i.e., first the summand~\eqref{Equ:ConcreteSummand} is expanded and afterwards the summation machinery is applied. More precisely, with the following function call we arrive at the coefficients $I_{-3}(n),\dots,I_1(n)$ in terms of harmonic sums and the Riemann Zeta values $z_i=\zeta(i)=\sum_{k=1}^{\infty}1/k^i$:

\begin{Cmma}
\CIn \!\!EvaluateMultiSum[f\!,\{\{i,0,k\},\{k,0,\infty\},\{j,0,m\},\{m,0,N\}\},\{n\},\{2\},\{\infty\},ExpandIn\hspace*{-0.14cm}\to\hspace*{-0.14cm}\{\ep,-3,1\}]\\ 
\COut \{-\frac{8}{3 (n+1)^2}, -\frac{8 S[1,n]}{3 (n+1)^2},
 -\frac{2 SS[1,n]^2}{3 (n+1)^2}-\frac{2 S[2,n]}{3 (n+1)^2}-\frac{\text{z}_ 2}{(n+1)^2}-\frac{8}{3 (n+1)^4},\newline
 -\frac{S[1,n]^3}{9 (n+1)^2}+\big(-\frac{S[2,n]}{(n+1)^2}-\frac{8}{3 (n+1)^4}\big) S[1,n]-\frac{\text{z}_ 2 S[1,n]}{(n+1)^2}-\frac{8 S[3,n]}{9 (n+1)^2}-\frac{2 S[2,1,n]}{3 (n+1)^2}-\frac{5 \text{z}_ 3}{3 (n+1)^2},\newline
 -\frac{S[1,n]^4}{72 (n+1)^2}+\big(-\frac{S[2,n]}{4 (n+1)^2}-\frac{2}{3 (n+1)^4}\big) S[1,n]^2+\big(\frac{S[2,n]}{(n+1)^3}+\frac{2 S[3,n]}{9 (n+1)^2}-\frac{S[2,1,n]}{(n+1)^2}\big) S[1,n]-\frac{3 S[2,n]^2}{8 (n+1)^2}-\frac{23 \text{z}_ 2^2}{16 (n+1)^2}-\frac{2 S[2,n]}{3 (n+1)^4}+\frac{S[3,n]}{(n+1)^3}+\frac{S[3,n]}{12 (n+1)^2}-\frac{S[2,1,n]}{(n+1)^3}-\frac{5 S[3,1,n]}{3 (n+1)^2}+\frac{5 S[2,1,1,n]}{3 (n+1)^2}+\big(-\frac{S[1,n]^2}{4 (n+1)^2}-\frac{S[2,n]}{4 (n+1)^2}-\frac{1}{(n+1)^4}\big) \text{z}_ 2+\big(\frac{2}{(n+1)^3}-\frac{5 S[1,n]}{3 (n+1)^2}\big) \text{z}_3-\frac{8}{3 (n+1)^6}\}\\
\end{Cmma}

\noindent We remark that during these calculations exceptional points at the summation borders are carefully treated (like, e.g., the point $i=0$ that does not hold for the sum representation~\eqref{Equ:4FoldEpFree}).

As demonstrated above, the definite summation spiral in Fig.~\ref{Fig:Spiral} (finding a recurrence, solving the recurrence, combining the recurrence using initial values\footnote{The initial values (e.g., $n=1,2,3$) can be obtained by the same method (with one parameter less); see~\cite{Schneider:07c}. As a consequence, we find indefinite nested product-sum expressions which 
usually simplify to multiple zeta values, infinite versions of $S$-sums or cyclotomic sums. Then techniques from \cite{Vermaseren:99,MZV,ABS:11,ABS:13} are heavily needed to rewrite the constants in this special form and to express them in terms of constants such that no further algebraic relations are known. In order to deal with such problems, Ablinger's \texttt{HarmonicSums} package is used.}) is applied iteratively from inside to outside, and the corresponding sums are transformed stepwise to indefinite nested product-sum expressions. 
For a description of the full method we refer to~\cite{Schneider:12a,Schneider:13b}. 
Here we want to stress the following aspect: For an arbitrary input, the method might fail. First, there might not exist a recurrence for a certain subproblem. However, for Feynman integrals as given in Section~\ref{Sec:Intro} the arising multi-sums have the appropriate shape to guarantee that the recurrence finder is always successful; this follows by ideas from~\cite{Wilf:92,AequalB,Wegschaider,AZ:06}. Here only time and space resources might be the bottleneck. Second, our recurrence solver might fail to find sufficiently many solutions in terms of indefinite nested product-sum expressions. And exactly here the miracle happens --at least for the classes of Feynman integrals that we considered so far: Problem~RS produces usually the full solution space for the occurring recurrences. Consequently, the solutions can be always combined to an alternative representation for the input sum. 
In summary, the presented method works very well for big classes of Feynman integrals.

\section{Crunching sums and mass production: the package \texttt{SumProduction}}\label{Sec:SumProduction}

So far we presented symbolic summation technologies and the related packages that enables one to simplify multi-sums to expressions in terms of indefinite nested product-sum expressions. For various situations, in particular for our 2-loop calculations~\cite{Schneider:08e} (based on a careful preparation of my cooperation partners) and case studies of massive 3-loop scalar ladder integrals~\cite{ABHKSW:12}, this toolbox was sufficient to perform the necessary calculations.

However, in 3-loop calculations for complete physical problems in QCD the number of the occurring sums grows substantially, i.e., several thousands, even up to several hundred thousands of sums have to be simplified. Typical examples are, e.g., the calculations of the first two complete Wilson coefficients $L_{qq,Q}^{PS}$ and $L_g^S$ for general values of the Mellin variable $n$; see~\cite{ABKSW:11}. Further examples are the current calculations of 3-loop graphs with two fermionic lines of equal mass and diagrams with two massive lines of different mass (for charm and bottom quarks); for examples see~\cite{DIS:13}. 

For all these problems the additional package \texttt{SumProduction} was heavily used to perform these large scale problems. Subsequently, the feature and usage of the package will be illustrated by the 3-loop corrections of $O(n_f T_F^2 C_{A,F})$ to the massive 
OMEs with local operator insertions on the gluonic lines, $A_{gq,Q}$ and $A_{gg,Q}$ 
at general values of the Mellin variable $n$~\cite{BHKS:13}. One of the larger expressions (actually, it is one of the smallest examples in comparison to the other examples mentioned above) was produced with the help of FORM~\cite{FORM,Vermaseren:13}. I.e., it is a 2 GB expression of 2419 multi--sums. Each of them can be treated by EvaluateMultiSum, like e.g.,

\begin{Cmma}\CMLabel{MMA:DoubeSumExp}
\CIn EvaluateMultiSum[\tfrac{\pi  2^{\ep+3} e^{-\frac{3 \gamma  \ep}{2}} (-1)^{j_1} (j_2+1) \Gamma (2-\ep) \Gamma \left(\frac{\ep}{2}+2\right) \Gamma \left(-\frac{3 \ep}{2}\right) \Gamma
   \left(-\frac{\ep}{2}+j_1+4\right) \Gamma (-j_1+n-2) \Gamma (\ep-j_1-j_2+n-5)}{(\ep-10) (\ep-8) (\ep-2) \ep \Gamma
   \left(\frac{5}{2}-\ep\right) \Gamma \left(\frac{\ep+5}{2}\right) \Gamma \left(\frac{\ep}{2}+n+1\right) \Gamma (-j_1-j_2+n-4)},\vspace*{0.1cm}\newline
\hspace*{2cm}	\{\{j_2,0,n-j_1-6\},\{j_1,0,n-5\}\},\{n\},\{5\},ExpandIn\to\{\ep,-3,-1\}]\vspace*{0.1cm}\\
\COut \big\{0,\frac{16 (-1)^n \big(3 n^2+12 n+11\big)}{135 (n+1) (n+2)^2 (n+3)^2}
-\frac{16 \big(n^8+6 n^7-6 n^6-80 n^5-81 n^4+178 n^3+274 n^2-4 n-96\big)}{45 (n-2) (n-1)^2 n^2 (n+1) (n+2)^2 (n+3)^2}\newline
\frac{16 \big(n^2-n-8\big)}{45 (n-1) n (n+2) (n+3)}S[1,n],-\frac{8 \big(n^2-n-8\big)}{45 (n-1) n (n+2) (n+3)}S[2,n]+\frac{2 (-1)^n (187 n+127) \big(3 n^2+12 n+11\big)c}{2025 (n+1)^2 (n+2)^2 (n+3)^2}\newline
+\big(\frac{2 \big(17 n^6-231 n^5+121 n^4+2063 n^3-1458 n^2-2432 n+960\big)}{675 (n-2) (n-1)^2 n^2 (n+1) (n+2) (n+3)}-\frac{16 (-1)^n \big(3 n^2+12 n+11\big)}{135 (n+1) (n+2)^2 (n+3)^2}\big) S[1,n]\newline
+\tfrac{2 \big(43 n^{12}+112 n^{11}+263 n^{10}-216 n^9-11309 n^8-16476 n^7+55837 n^6+78164 n^5-95178 n^4-116688 n^3+51784 n^2+30624 n-23040\big)}{675 (n-2)^2 (n-1)^3 n^3 (n+1)^2 (n+2)^2 (n+3)^2}\big\}\\
\end{Cmma}

\noindent For details on the calculation steps for this particular sum we refer to~\cite{Schneider:12a}. Similarly, all the other sums could be treated step by step with a lot of computer resources. However, as worked out in~\cite{Schneider:12a} we can do it much better by using the toolbox of the package 

\begin{Cmma}
\CIn << SumProduction.m \\
\CPrint SumProduction - A summation package by Carsten Schneider \copyright\ RISC-Linz\\
\end{Cmma}

\vspace*{-0.6cm}

\subsection{Reduction to master sums.} 
First, we reduce the 2 GByte expression (stored in \texttt{expr} and being valid for $n\geq6$) to master sums (resp.\ key sums) with the function call
\begin{Cmma}
\CIn compactExpr=ReduceMultiSums[expr,\{n\},\{6\},\{\infty\}];\\
\end{Cmma}
\noindent The reduced expression \texttt{compactExpr} is only 7.6 MByte large and it required 6 hours and 53 minutes to obtain this reduction.

\medskip

\noindent\MyFrame{\noindent\textbf{Problem RMS: ReduceMultiSums.}
\textit{Given} a linear combination of definite multi-sums\footnote{I.e., the sums can be of the form~\eqref{Eq:GenericMultiSum} like in our concrete example~\CmyIn{\ref{MMA:DoubeSumExp}}, or the summands might also involve, e.g., harmonic sums in the numerators; cf.~\eqref{Equ:MultiSumsOverIndefNested}. In addition, further regulators (like $\ep$) might be involved.} over indefinite nested product-sum expressions as given in~\eqref{Equ:EMSInputSum}.
\textit{Compactify} the expression, i.e., express the summands with objects such that no algebraic relations remain. In particular, synchronize the summation bounds and merge the sums to  master sums.
}

\medskip

\noindent In this routine the merging can be done in different ways. By default, the summands are tried to be given in the form $r\,t_1^{m_1}\dots t_r^{m_r}$ where the $t_i$ are indefinite nested sums or products, $m_i\in\set Z$ and $r$ is a rational function where the numerator and denominator are co-prime.

In our concrete example the 2419 sums are synchronized w.r.t.\ the occurring summation ranges (taking for each class the maximum of the lower bounds and the minimum of the upper bounds). As result, we obtained only 4 sums with synchronized ranges
$$\sum_{i_2=5}^{n-5}\sum_{i_1=0}^{i_2}h_1(\varepsilon,n,i_2,i_1),\;\;\;\sum_{i_2=0}^{n-5}\sum_{i_1=0}^{n-i_2-5}h_2(\varepsilon,n,i_2,i_1),\;\;\;\sum_{i_1=5}^{n-5}h_3(\varepsilon,n,i_1),\;\;\;\sum_{i_1=0}^{\infty}h_4(\varepsilon,n,i_1)$$
plus a large term free of summation quantifiers.
Next, all the occurring Pochhammer symbols, factorials/$\Gamma$-functions, and binomials are written in a basis of algebraically independent objects plus the extra object $(-1)^n$ (if necessary); for details see Problem~EAR in Subsection~\ref{Sec:IndefSum}. 
Finally, the expressions are split further to get the form $\sum h(n,(i_2,)i_1,\ep)*r(n,(i_2,)i_1,\ep)$ or $h(n,\ep)*r(n,\ep)$ where $h$ stands for a (proper) hypergeometric term in $n$ (and $i_1,i_2$), i.e., being a product of binomials/factorials/Pochhammers in the numerator and denominator, and $r(n,(i_2,i_1),\ep)$ being a rational function in $n,\ep$ (and $i_1,i_2)$; note that $r$ might fill several pages. As final result we obtain an expression with only 29 sums and 15 terms being free of sums.

\subsection{Automatic computation of the $\ep$--expansions (in parallel)} 
Finally, the sums are simplified with \texttt{EvaluateMultiSums}. In particular, the coefficients of the required Laurent series expansion can be derived. In order to perform this calculation automatically, the following function call can be applied:

\begin{Cmma}
\CIn ProcessEachSum[compactExpr,\{n\},\{6\},\{\infty\},ExpandIn\to\{\ep,-3,0\}]\\
\end{Cmma}

\noindent It sequentially applies \texttt{EvaluateMultiSum} with the corresponding input parameters to the occurring multi-sums in \texttt{compactExpr}. In our concrete example this step took in total 2 hours and 35 minutes.

Often the evaluation of one sum (in particular, for triple and quadruple sums) takes several hours. In order to utilize the benefit of the available computers, we emphasize that this function can be executed simultaneously with different Mathematica kernels, in particular, on different machines within a network.
Here the following mechanism is applied. Internally, the function takes the first multi-sum and generates a file with the name SUM1. If the result is computed, the file is updated with the result. Then the routine continues with the second sum provided the file SUM2 is not existent on the hard disk. In this way, ProcessEachSum can be executed in parallel for mass productions.

\subsection{Combination to the final result} Finally, the result of the sums (or the expansion of the sums) are read from the hard disk and are summed up to the final result. 
In particular, the expressions are reduced further by applying Problem~EAR given in Subsection~\ref{Sec:IndefSum} to the occurring sums and products. As shortcut also the available algebraic relations for harmonic sums and their generalized versions~\cite{Bluemlein:04,ABS:11,ABS:13,ABS:13b} are utilized. This last step can be carried out with the following function call

\begin{Cmma}
\CIn result=CombineExpression[compactExpr,\{n\},\{6\},\{\infty\}];\\
\end{Cmma}

\noindent This calculation took about 21 seconds. The final result can be expressed in terms of 
$\zeta_2$, $\zeta_3$, $(-1)^n$, $S_1(n)$, $S_2(n)$, $S_3(n)$, $S_{2,1}(n)$, $S_{3,1}(n)$, $S_{2,1,1}(n)$
and requires about 100 KByte memory.
In summary, the total calculation took around 9 hours and 30 minutes.

\section{Conclusion}\label{Sec:Conclusion}

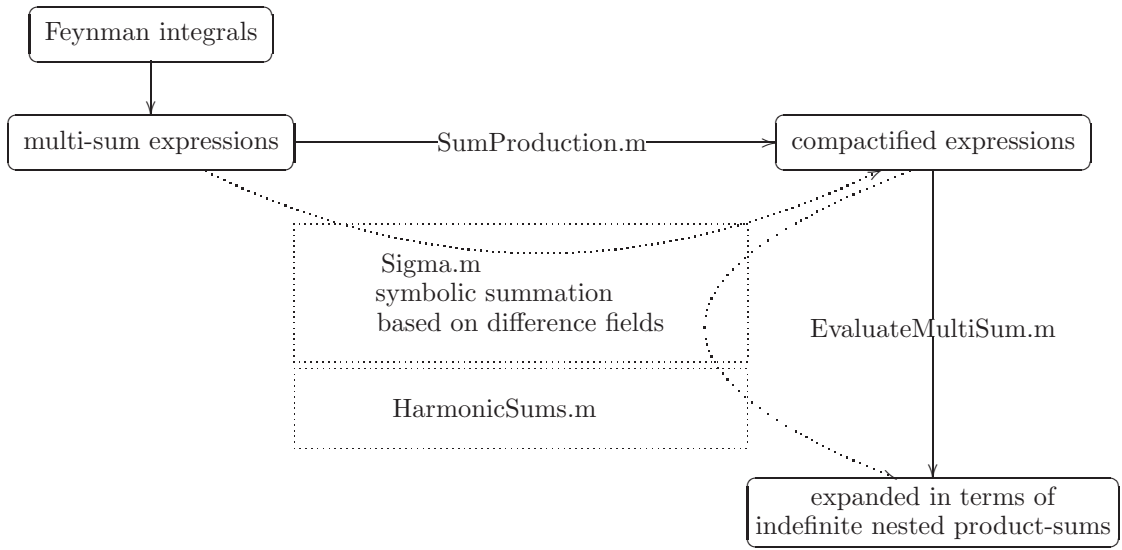
\begin{figure}
  \centering
\small
\xymatrix@C=0cm@R=.1cm{
*+<0.2cm,0.2cm>+[F-:<3pt>]\txt{Feynman integrals}\ar[ddd]\\
\\\\
*+<0.2cm,0.2cm>+[F-:<3pt>]\txt{multi-sum expressions}\ar[rr]|{\txt{SumProduction.m}}\ar@{.>}@/_3.5pc/[rr] &&*+<0.2cm,0.2cm>+[F-:<3pt>]\txt{compactified expressions}\ar[dddddd]|{\txt{EvaluateMultiSum.m}}\ar@<0.8cm>@{.>}@/_9.1pc/[dddddd]] \\
\\\\
&*+<0.cm,0.6cm>+[F..]\txt{
\hspace*{-2.5cm}Sigma.m\\
\hspace*{-0.7cm}symbolic summation\\
\hspace*{1cm}based on difference fields\hspace*{1cm}}\\
&*+<3.8cm,.6cm>+[F..]\txt{\hspace*{-0.7cm}HarmonicSums.m}\hspace*{2cm}&\\
\\
&& *+[F-:<3pt>]\txt{expanded in terms of\\ indefinite nested product-sums}\\
}
\normalsize
\caption{The packages in interaction\label{Fig:Packages}}
\end{figure}

Summarizing, the whole interaction of the presented packages can be visualized in Fig.~\ref{Fig:Packages}. 
Here the Feynman integrals are transformed to huge multi-sum expressions using mostly the computer algebra system FORM. These expressions are then loaded into the computer algebra system Mathematica and our machinery is activated. Using the package \texttt{SumProduction} the multi-sum expressions are crunched to expressions in terms of master sums/key sums. In addition, the package supports the user to perform the simplification completely automatically on distributed systems.
In order to perform these simplifications, in particular to calculate the coefficients of the Laurent series expansion, the package \texttt{EvaluateMultiSums} is called accordingly. 
This summation technology relies heavily on the summation toolkit of the \texttt{Sigma} package that is based on difference field theory.
In addition, special function algorithms are needed if infinite summations arise in the given expressions. Here we rely on Ablinger's HarmonicSums package~\cite{ABS:11,Ablinger:12,ABS:13,ABS:13b} that utilizes and generalizes ideas of~\cite{Vermaseren:99,Remiddi2000,Bluemlein:09,MZV}.

All the packages and underlying algorithms are steadily extended, improved and optimized to deal with more and more complicated Feynman integrals.

\ack
This work is supported by the Austrian Science Fund (FWF) grants P20347-N18 and SFB F50 (F5009-N15) and
by the EU Network {\sf LHCPhenoNet} PITN-GA-2010-264564.

\section*{References}

\begin{thebibliography}{10}
\expandafter\ifx\csname url\endcsname\relax
  \def\url#1{{\tt #1}}\fi
\expandafter\ifx\csname urlprefix\endcsname\relax\def\urlprefix{URL }\fi
\providecommand{\eprint}[2][]{\url{#2}}

\bibitem{Karr:81}
Karr M 1981 {\em J.~ACM\/} {\bf 28} 305--350

\bibitem{Schneider:05f}
Schneider C 2005 {\em Proc. ISSAC'05\/} ed Kauers M (ACM) pp 285--292

\bibitem{Schneider:08c}
Schneider C 2008 {\em J. Symbolic Comput.\/} {\bf 43} 611--644
  [arXiv:0808.2543v1]

\bibitem{Schneider:10b}
Schneider C 2010 {\em {Motives, Quantum Field Theory, and Pseudodifferential
  Operators}\/} ({\em Clay Mathematics Proceedings\/} vol~12) ed Carey A,
  Ellwood D, Paycha S and Rosenberg S (Amer. Math. Soc) pp 285--308
  arXiv:0808.2543

\bibitem{Schneider:04a}
Schneider C 2004 {\em Proc. ISSAC'04\/} ed Gutierrez J (ACM Press) pp 282--289

\bibitem{Schneider:07d}
Schneider C 2007 {\em J. Algebra Appl.\/} {\bf 6} 415--441

\bibitem{Schneider:10c}
Schneider C 2010 {\em Ann. Comb.\/} {\bf 14} 533--552 [arXiv:0808.2596]

\bibitem{Schneider:13a}
Schneider C 2013 {\em {Computer Algebra in Quantum Field Theory: Integration,
  Summation and Special Functions}\/} Texts and Monographs in Symbolic
  Computation ed Schneider C and Bl\"umlein J (Springer) pp 325--360
  arXiv:1304.4134 [cs.SC]

\bibitem{Schneider:01}
Schneider C 2011 {\em Symbolic Summation in Difference Fields\/} Ph.D. thesis
  RISC, Johannes Kepler University, Linz technical report 01-17

\bibitem{Schneider:04c}
Schneider C 2004 {\em Discrete Math. Theor. Comput. Sci.\/} {\bf 6} 365--386

\bibitem{Schneider:07a}
Schneider C 2007 {\em S\'em.~Lothar. Combin.\/} {\bf 56} 1--36 article B56b

\bibitem{Zeilberger:91}
Zeilberger D 1991 {\em J.~Symbolic Comput.\/} {\bf 11} 195--204

\bibitem{Petkov:92}
Petkov{\v s}ek M 1992 {\em J.~Symbolic Comput.\/} {\bf 14} 243--264

\bibitem{AequalB}
Petkov{\v s}ek M, Wilf H~S and Zeilberger D 1996 {\em $A=B$\/} (Wellesley, MA:
  A. K. Peters)

\bibitem{Schneider:03}
Paule P and Schneider C 2003 {\em Adv. in Appl. Math.\/} {\bf 31} 359--378

\bibitem{APS:05}
Andrews G, Paule P and Schneider C 2005 {\em Advances in Applied Math.\/} {\bf
  34} 709--739

\bibitem{Schneider:06c}
Driver K, Prodinger H, Schneider C and Weideman J~A~C 2006 {\em Ramanujan~J.\/}
  {\bf 12} 299--314

\bibitem{Schneider:09a}
Osburn R and Schneider C 2009 {\em Math. Comp.\/} {\bf 78} 275--292
  arXiv:math/0610281 [math.NT]

\bibitem{ABRS:12}
Ablinger J, Bl\"umlein J, Round M and Schneider C 2012 {\em {Loops and Legs in
  Quantum Field Theory 2012}\/} PoS(2012)50 pp 1--14 arXiv:1210.1685 [cs.SC]

\bibitem{Schneider:12a}
Bl\"umlein J, Hasselhuhn A and Schneider C 2012 {\em {Proceedings of RADCOR
  2011}\/} vol PoS(RADCOR2011)32 pp 1--9 arXiv:1202.4303 [math-ph]

\bibitem{Schneider:10d}
Ablinger J, Bl\"umlein J, Klein S and Schneider C 2010 {\em Nucl. Phys. B
  (Proc. Suppl.)\/} {\bf 205-206} 110--115 arXiv::1006.4797 [math-ph]

\bibitem{Bluemlein:09}
Bl\"umlein J 2009 {\em Comput. Phys. Commun.\/} {\bf 180} [arXiv:0901.3106
  [hep-ph]]

\bibitem{BKSF:12}
Bl\"umlein J, Klein S, Schneider C and Stan F 2012 {\em J. Symbolic Comput.\/}
  {\bf 47} 1267--1289 arXiv:1011.2656 [cs.SC]

\bibitem{BW:10}
Bogner C and Weinzierl S 2010 {\em Int.\ J.\ Mod.\ Phys.\ A\/} {\bf 25}
  2585--2618 [arXiv:1002.3458 [hep-ph]]

\bibitem{Weinzierl:13}
Weinzierl S 2013 {\em {Computer Algebra in Quantum Field Theory: Integration,
  Summation and Special Functions}\/} Texts and Monographs in Symbolic
  Computation ed Schneider C and Bl\"umlein J (Springer) pp 381--406

\bibitem{Ablinger:12}
Ablinger J 2012 {\em Computer Algebra Algorithms for Special Functions in
  Particle Physics\/} Ph.D. thesis J. Kepler University Linz

\bibitem{AZ:06}
Apagodu M and Zeilberger D 2006 {\em Adv. Appl. Math.\/} {\bf 37} 139--152

\bibitem{Brown:09}
Brown F 2009 {\em Math. Phys.\/} {\bf 287} 925--958

\bibitem{ABHKSW:12}
Ablinger J, Bl\"umlein J, Hasselhuhn A, Klein S, Schneider C and Wissbrock F
  2012 {\em Nuclear Physics B\/} {\bf 864} 52--84 arXiv:1206.2252v1 [hep-ph]

\bibitem{Wilf:92}
Wilf H and Zeilberger D 1992 {\em Invent. Math.\/} {\bf 108} 575--633

\bibitem{Wegschaider}
Wegschaider K 1997 {\em Computer generated proofs of binomial multi-sum
  identities\/} Master's thesis RISC, J. Kepler University

\bibitem{Schneider:05d}
Schneider C 2005 {\em Adv. in Appl. Math.\/} {\bf 34} 740--767

\bibitem{Schneider:05a}
Schneider C 2005 {\em J. Differ. Equations Appl.\/} {\bf 11} 799--821

\bibitem{Chyzak:00}
Chyzak F 2000 {\em Discrete Math.\/} {\bf 217} 115--134

\bibitem{Vermaseren:99}
Vermaseren J~A~M 1999 {\em Int. J.~Mod. Phys.\/} {\bf A14} 2037--2976
  arXiv:hep-ph/9806280

\bibitem{Bluemlein:99}
Bl\"umlein J and Kurth S 1999 {\em Phys. Rev.\/} {\bf D60} arXiv:hep-ph/9810241

\bibitem{Moch:02}
Moch S~O, Uwer P and Weinzierl S 2002 {\em J. Math. Phys.\/} {\bf 43}
  3363--3386

\bibitem{ABS:13}
Ablinger J, Bl\"umlein J and Schneider C 2013 {\em J. Math. Phys.\/} {\bf 54}
  1--74 arXiv:1302.0378 [math-ph]

\bibitem{ABS:11}
Ablinger J, Bl\"umlein J and Schneider C 2011 {\em J. Math. Phys.\/} {\bf 52}
  1--52 [arXiv:1007.0375 [hep-ph]]

\bibitem{DIS:13}
Ablinger J, Bl\"umlein J, Freitas A~D, Hasselhuhn A, von Manteuffel A, Raab C,
  Round M, Schneider C and Wissbrock F 2013 {\em {XXI International Workshop on
  Deep-Inelastic Scattering and Related Subjects - DIS2013}\/} arXiv:1307.7548
  [hep-ph]

\bibitem{AB:2013}
Ablinger J and Bl\"umlein J 2013 {\em Computer Algebra in Quantum Field Theory:
  Integration, Summation and Special Functions\/} Texts and Monographs in
  Symbolic Computation ed Schneider C and Bl\"umlein J (Springer) pp 1--32

\bibitem{Weinzierl:04}
Weinzierl S 2004 {\em J. Math. Phys.\/} {\bf 45} 2656--2673
  arXiv:hep-ph/0402131

\bibitem{ACAT:07}
Bierenbaum I, Bl\"umlein J, Klein S and Schneider C 2007 {\em {Proc. ACAT
  2007}\/} vol PoS(ACAT)082 pp 1--15 arXiv:0707.4659 [math-ph]

\bibitem{Schneider:08e}
Bierenbaum I, Bl{\"u}mlein J, Klein S and Schneider C 2008 {\em Nucl.Phys. B\/}
  {\bf 803} 1--41 [arXiv:hep-ph/0803.0273]

\bibitem{ACAT:08}
Bl\"umlein J, Kauers M, Klein S and Schneider C 2008 {\em {Proc. of ACAT
  2008}\/} vol PoS(ACAT08)106 pp 1--7 arXiv:0902.4095 [hep-ph]

\bibitem{BKKS:09}
Bl{\"u}mlein J, Kauers M, Klein S and Schneider C 2009 {\em Comput. Phys.
  Commun.\/} {\bf 180} 2143--2165 arXiv:0902.4091 [hep-ph]

\bibitem{LLPhys:10}
Ablinger J, Bierenbaum I, Bl\"umlein J, Hasselhuhn A, Klein S, Schneider C and
  Wissbrock F 2010 {\em Nucl. Phys. B (Proc. Suppl.)\/} {\bf 205-206} 242--249
  arXiv::1007.0375 [hep-ph]

\bibitem{DIS:11}
Ablinger J, Bl\"umlein J, Klein S, Schneider C and Wissbrock F 2011 {\em {19th
  International Workshop On Deep-Inelastic Scattering And Related Subjects (DIS
  2011)}\/} (American Institute of Physics (AIP)) arXiv:1106.5937 [hep-ph]

\bibitem{ABKSW:11}
Ablinger J, Bl\"umlein J, Klein S, Schneider C and Wissbrock F 2011 {\em Nucl.
  Phys. {\bf B}\/} {\bf 844} 26--54 arXiv:1008.3347 [hep-ph]

\bibitem{LLPhys:12}
Ablinger J, Bl\"umlein J, Freitas A~D, Hasselhuhn A, Klein S, Raab C, Round M,
  Schneider C and Wissbrock F 2012 {\em {Proc. Loops and Legs in Quantum Field
  Theory 2012}\/} PoS(LL2012)033 pp 1--12 arXiv:1212.6823 [hep-ph]

\bibitem{DIS:12}
Ablinger J, Bl\"umlein J, Freitas A~D, Hasselhuhn A, Klein S, Schneider C and
  Wissbrock F 2012 {\em {Proceedings of the 36th International Conference on
  High Energy Physics}\/} vol PoS(ICHEP2012)270 pp 1--9 arXiv:1212.5950
  [hep-ph]

\bibitem{BHKS:13}
Bl\"umlein J, Hasselhuhn A, Klein S and Schneider C 2013 {\em Nuclear Physics
  B\/} {\bf 866} 196--211 arXiv:1205.4184 [hep-ph]

\bibitem{Gosper:78}
Gosper R~W 1978 {\em Proc. Nat. Acad. Sci. U.S.A.\/} {\bf 75} 40--42

\bibitem{Zeilberger:90a}
Zeilberger D 1990 {\em J.~Comput. Appl. Math.\/} {\bf 32} 321--368

\bibitem{KP:11}
Kauers M and Paule P 2011 {\em The concrete tetrahedron\/} Texts and Monographs
  in Symbolic Computation (Springer)

\bibitem{Koutschan:13}
Koutschan C 2013 {\em {Computer Algebra in Quantum Field Theory: Integration,
  Summation and Special Functions}\/} Texts and Monographs in Symbolic
  Computation ed Schneider C and Bl\"umlein J (Springer) pp 171--194
  arXiv:1307.4554 [cs.SC]

\bibitem{Kauers:07}
Kauers M 2007 {\em J. of Symbolic Comput.\/} {\bf 42} 948--970

\bibitem{Schneider:13b}
Schneider C 2013 {\em {arXiv:1307.7887 [cs.SC]}\/}

\bibitem{Karr:85}
Karr M 1985 {\em J.~Symbolic Comput.\/} {\bf 1} 303--315

\bibitem{Bluemlein:04}
Bl{\"u}mlein J 2004 {\em Comput. Phys. Commun.\/} {\bf 159} 19--54
  [arXiv:hep-ph/0311046]

\bibitem{ABS:13b}
Ablinger J, Bl\"umlein J and Schneider C 2013 {\em In preparation: Structural
  Relations of Harmonic Sums\/}

\bibitem{Hoffman:00}
Hoffman M 2000 {\em J. Algebraic Combin.\/} {\bf 11} 49--68

\bibitem{Schneider:10a}
Schneider C 2010 {\em Appl. Algebra Engrg. Comm. Comput.\/} {\bf 21} 1--32

\bibitem{Schneider:05c}
Schneider C 2005 {\em Ann. Comb.\/} {\bf 9} 75--99

\bibitem{Petkov:10}
Abramov S~A and Petkov{\v{s}}ek M 2010 {\em J. Symbolic Comput.\/} {\bf 45}
  684--708

\bibitem{ABPS:13}
Abramov S~A, Bronstein M, Petkov\v{s}ek M and Schneider C 2013 {\em In
  preparation\/}

\bibitem{Bron:00}
Bronstein M 2000 {\em J.~Symbolic Comput.\/} {\bf 29} 841--877

\bibitem{Abramov:94}
Abramov S~A and Petkov{\v s}ek M 1994 {\em Proc. ISSAC'94\/} ed von~zur Gathen
  J (ACM Press) pp 169--174

\bibitem{Singer:99}
Hendriks P~A and Singer M~F 1999 {\em J.~Symbolic Comput.\/} {\bf 27} 239--259

\bibitem{Petkov:2013}
Petkov{\v s}ek M and Zakraj{\v s}ek H 2013 {\em Computer Algebra in Quantum
  Field Theory: Integration, Summation and Special Functions\/} Texts and
  Monographs in Symbolic Computation ed Schneider C and Bl\"umlein J (Springer)
  pp 259--284

\bibitem{Moch:04}
Moch S~O, Vermaseren J~A~M and Vogt A 2004 {\em Nucl. Phys. B\/} {\bf 688} 101
  arXiv:hep-ph/0403192v1

\bibitem{Guessing:09}
Kauers M 2009 {Guessing Handbook} Tech. Rep. 09-07 RISC Report Series,
  University of Linz, Austria

\bibitem{Schneider:07c}
Pemantle R and Schneider C 2007 {\em Amer. Math. Monthly\/} {\bf 114} 344--350

\bibitem{MZV}
Bl\"umlein J, Broadhurst D~J and Vermaseren J~A~M 2010 {\em Comput.\ Phys.\
  Commun.\/} {\bf 181} 582--625 [arXiv:0907.2557 [math-ph]]

\bibitem{FORM}
Vermaseren J~A~M 2000 {\em arXiv:math-ph/0010025\/}

\bibitem{Vermaseren:13}
Vermaseren J~A~M 2013 {\em {Computer Algebra in Quantum Field Theory:
  Integration, Summation and Special Functions}\/} Texts and Monographs in
  Symbolic Computation ed Schneider C and Bl\"umlein J (Springer) pp 361--379

\bibitem{Remiddi2000}
Remiddi E and Vermaseren J~A~M 2000 {\em Int. J. Mod. Phys.\/} {\bf A 15} 725
  arXiv:hep-ph/9905237v1

\end{thebibliography}

\providecommand{\newblock}{}

\end{document}